\documentclass[fleqn,usenatbib]{mnras}

\usepackage{newtxtext,newtxmath}

\usepackage[T1]{fontenc}

\DeclareRobustCommand{\VAN}[3]{#2}
\let\VANthebibliography\thebibliography
\def\thebibliography{\DeclareRobustCommand{\VAN}[3]{##3}\VANthebibliography}

\usepackage{graphicx}	
\usepackage{amsmath}	

\title[I'm in AGNi]{I'm in AGNi: A new standard for AGN pluralisation}

\author[A. D.~Gow \emph{et al}.]{
Andrew D.~Gow$^{1}$\thanks{E-mail: \href{mailto:andrew.gow@port.ac.uk}{andrew.gow@port.ac.uk}},
Peter Clark$^{1}$
and Dan Rycanowski$^{1}$
\\
$^{1}$Institute of Cosmology \& Gravitation, University of Portsmouth, Dennis Sciama Building, Burnaby Road, Portsmouth, PO1 3FX, United Kingdom
}

\date{Accepted 01 April 2024. Received 01 April 2024; in original form 01 April 2024}

\pubyear{2024}

\begin{document}
\label{firstpage}
\pagerange{\pageref{firstpage}--\pageref{lastpage}}
\maketitle

\begin{abstract}
We present a new standard acronym for Active Galactic Nuclei, finally settling the argument of AGN \textit{vs.}~AGNs. Our new standard is not only etymologically superior (following the consensus set by SNe), but also boasts other linguistic opportunities, connecting strongly with relevant theology and streamlining descriptions of AGN properties.
\end{abstract}




\section{Introduction}

The center of most galaxies hosts a supermassive black hole (SMBH) which can display a wide range of accretion rates, from essentially zero to values up to a solar mass per day~\citep{Wolf:2024_Accretion}. The accretion of material onto a SMBH via an accretion disk can produce high luminosities through frictional heating of material generating high temperature blackbody emission, with this emission in some cases approaching the Eddington Luminosity. The specific threshold for when a galaxy's SMBH, and thus nucleus, is considered to be `active' is not concretely established though luminosity and central SMBH mass cuts are used to divide these galaxies from their more sedate compatriots. Such a nucleus is termed an Active Galactic Nucleus or AGN.

The presence of an AGN can dominate the spectral properties of the host galaxy, with the accreting SMBH contributing more light than the stellar population of the galaxy as a whole, despite occupying only a tiny portion of its overall volume. The specifics of these effects are wide ranging with some Active Galactic Nuclei displaying extremely broad spectroscopic emission features, others showing only narrow (but strong) emission lines. A subset of the population are very luminous in the radio (radio-loud) whilst others emit only faintly in this regime (radio-quiet). This range in properties posed a significant challenge in the the study of these objects, though these apparent dichotomies have been resolved through the development of the unified model of AGN physics. Detailed modeling revealed that the full range of displayed properties can indeed be attributed to accretion onto SMBHs when the structure of the local environment, line-of-sight effects from differential obscuration and the presence (or absence) of jets are considered, see~\citet{Netzer:2015_Unified} for a recent detailed review.

Whilst the unified model of AGN physics is widely accepted, and can be seen as as a beacon of unity within the astrophysics community, the choice of initialism when referring to Active Galactic Nuclei in plurality is far less harmonious. The two options currently in use are AGN~\citep{Arons:1975_AGN,Kauffmann:2003_AGN,Svoboda:2017_AGN,Mountrichas:2024_AGN,Clark:2024_AGN}, where the singular versus plural cases are distinguished via context, and AGNs~\citep{Yee:1980_AGNs,Kewley:2001_AGNs,Jiang:2016_AGNs,Ralowski:2024_AGNs,Clark:2024_AGNs}, with many examples of both appearing throughout the literature. This lack of standardisation is an unnecessary waste of time, and can potentially cause confusion when reading papers about Active Galactic Nuclei. Therefore, we believe a new standard is required.

\section{A new plural acronym}

To avoid any future disagreement about the correct way to pluralise this initialism, we introduce a new option which is superior to both previous choices, which we believe should become the standard case used in all literature. We suggest that the correct way to pluralise AGN is AGNi. This is etymologically stronger than either of the other choices, connecting to the Latin root of the word \textit{nucleus}, and follows the consensus set by other similar initialisms \textit{e.g.}~supernovae (SNe), (proto-)planetary nebulae ((P)PNe), and pulsar wind nebulae (PWNe).

Not only is this choice etymologically superior, but it allows for an increase in speed when discussing these objects. While AGN is an initialism, where the letters are read out individually, AGNi can be promoted to a full acronym, pronounced as a single word and thus boasting a significant 33.3 per cent increase in speaking efficiency---a metric widely utilised in the yapping community. This additionally provides many linguistic opportunities (puns) frequently used by the astronomy community. The simplest of these is the similarity to the word \textit{agony}, as used in the title of this paper. This extends to concepts such as an \textit{AGNi aunt} (a galaxy that displayed AGN activity in a past phase of evolution but is not currently active e.g., IC~2497~\cite{Lintott:2009_Voorwerp}), and \textit{AGNiser}~\citep{Agonizer:MA,Mirror:MA}  (bright AGNi that agonise astronomers by saturating all of their images). Additionally, any notable survey of AGNi led by a Swedish institution would of course be named \textit{AGNitha}, in honour of Agnetha Fältskog from ABBA.

\begin{figure}
\centering
\includegraphics[width=0.22\textwidth]{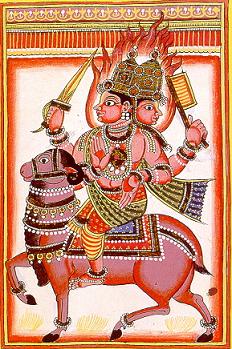}
\caption{Illustration of the Hindu god Agni~\citep{Agni:Wikimedia}.}
\label{fig:Agni}
\end{figure}

This acronym also provides a convenient shorthand for discourse in the field. For example the brightness across multiple AGNi would be simply the apparent or absolute \textit{mAGNitude}, and members of the community who are unsure of the legitimate status of AGNi would be \textit{AGNistic}. However, the benefits of this acronym do not stop with the ability to make puns. There is also a strong connection to theology through the Hindu god \textit{Agni}~\citep{Olivelle:1998_Agni,Lochtefeld:2002_Agni,Jamison:2014_Agni} (see fig.~\ref{fig:Agni}). \textit{Agni} is the god of fire (the name is connected to the English word \textit{ignite}), providing clear parallels with the high energy nature of AGNi.

\begin{figure*}
\centering
\includegraphics[width=0.5\textwidth]{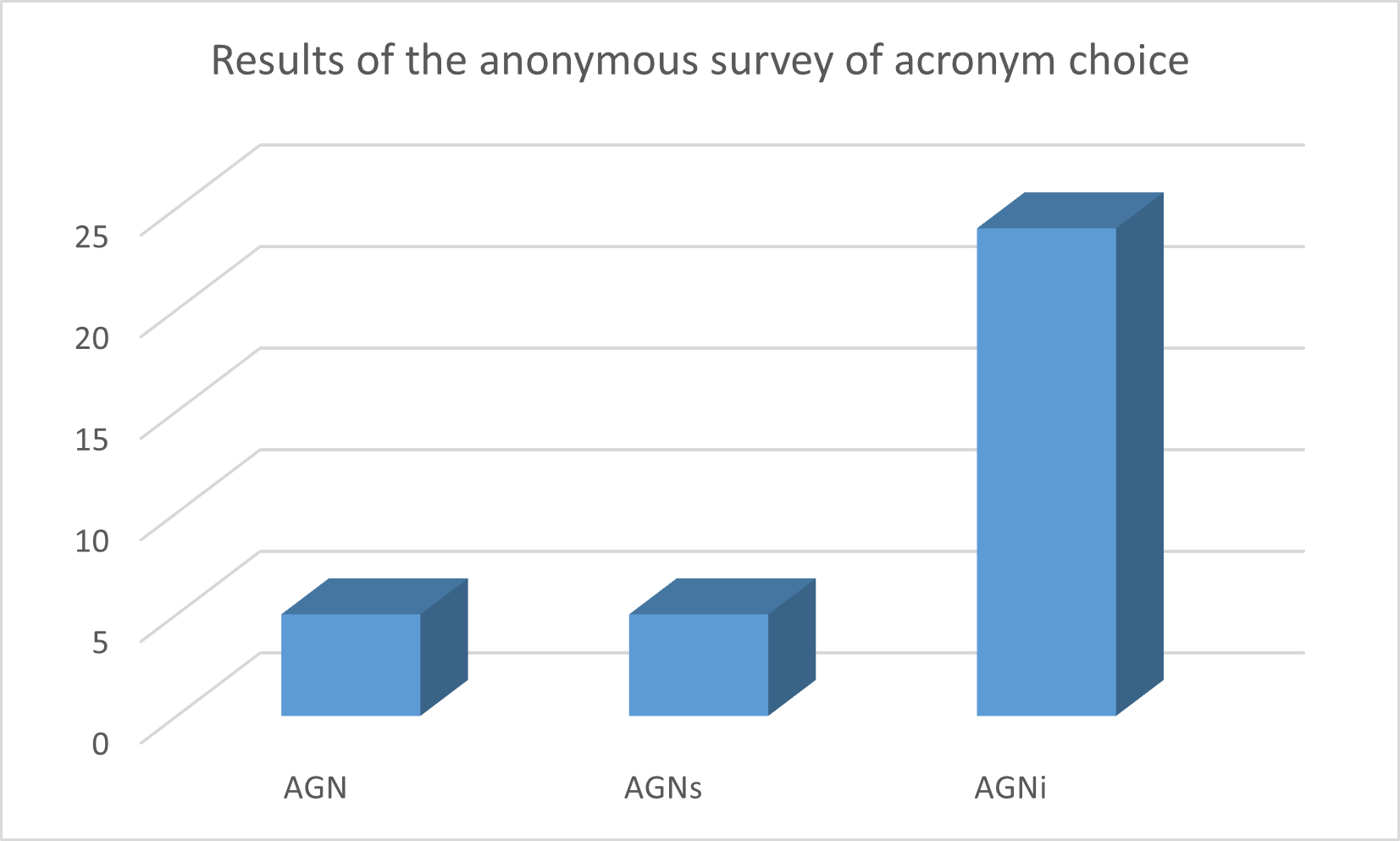}\includegraphics[width=0.5\textwidth]{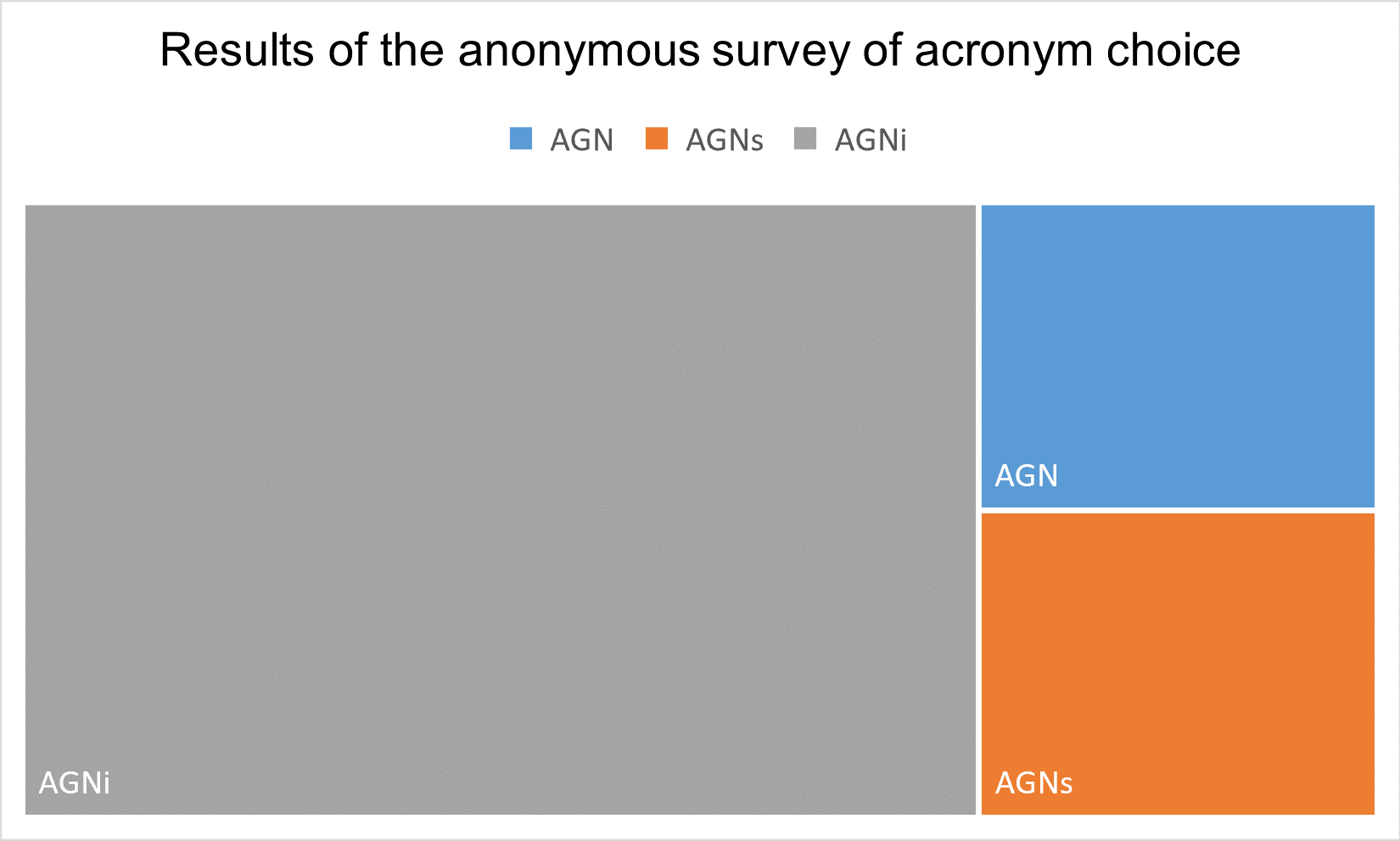}

\includegraphics[width=0.5\textwidth]{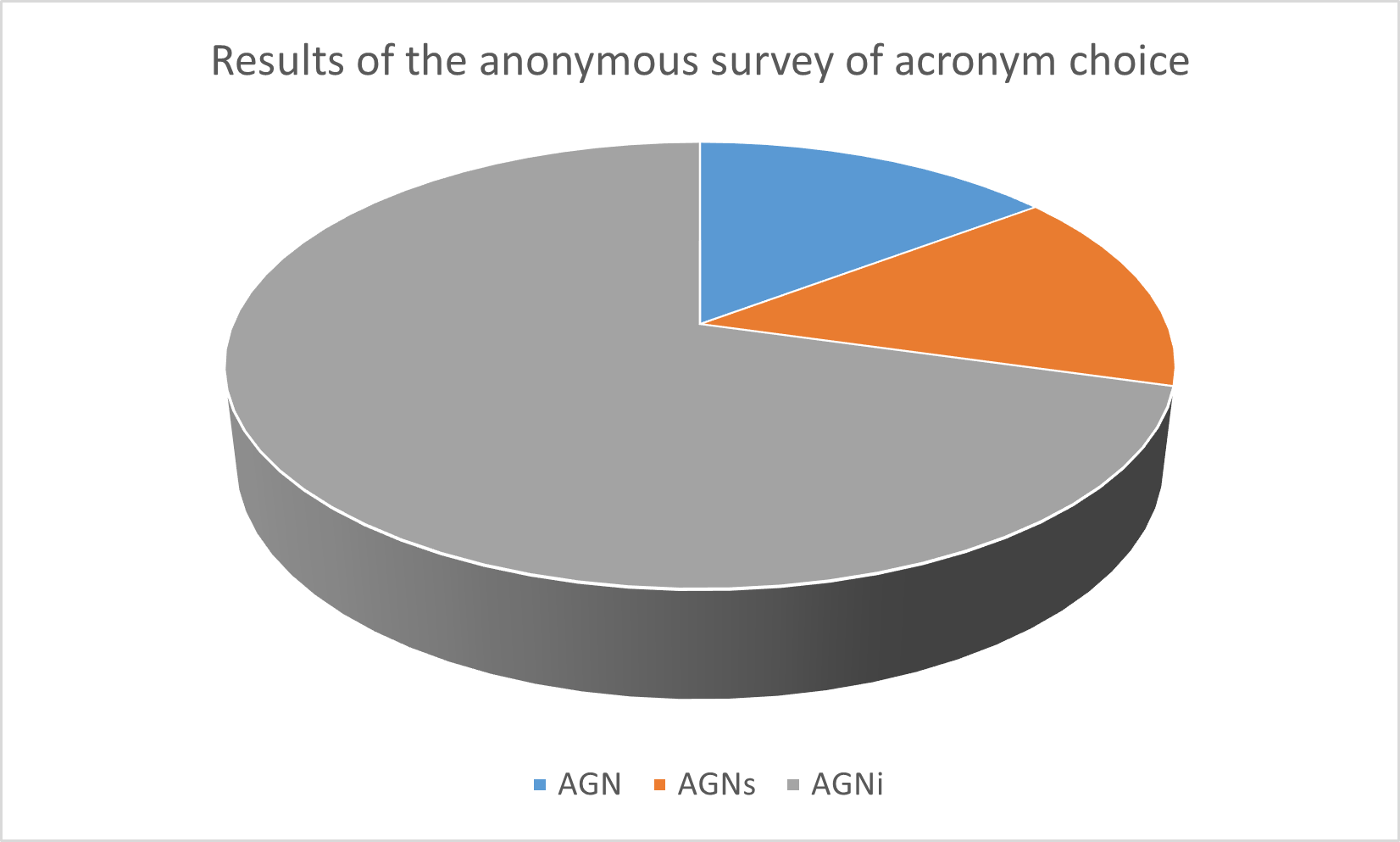}\includegraphics[width=0.5\textwidth]{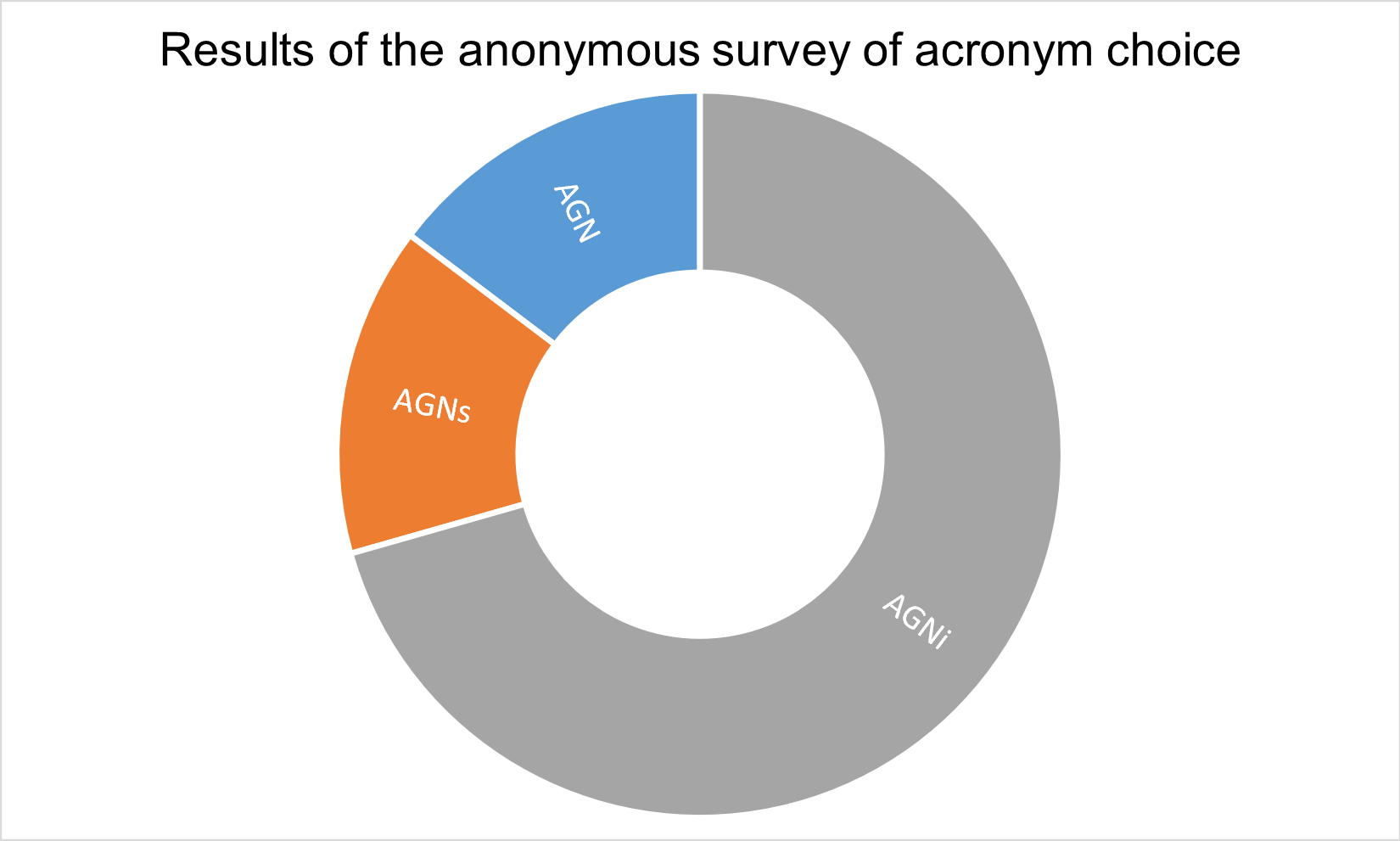}

\caption{Results of the anonymous survey of acronym choice.}
\label{fig:Survey-results}
\end{figure*}

\begin{figure*}
\centering
\includegraphics[width=\textwidth]{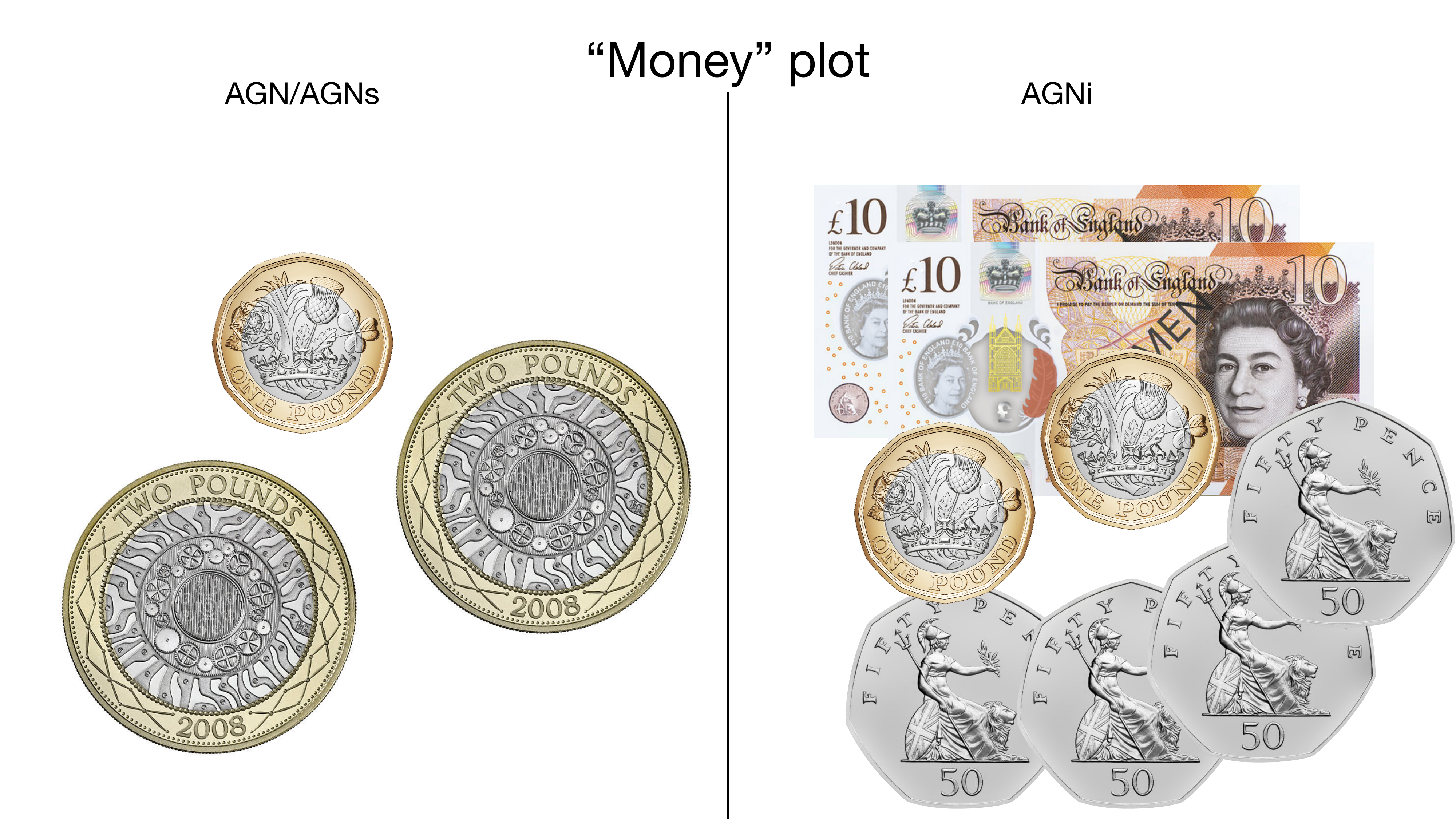}
\caption{Results of the anonymous survey of acronym choice.}
\label{fig:Money-plot}
\end{figure*}

However, for this acronym to be adopted, it is clear that there must be significant support from the scientific community. To this end, we have carried out an anonymous (pay no attention to any names in the acknowledgements) survey of astronomers and cosmologists, the results of which are displayed in table~\ref{tab:Survey-results} and figs.~\ref{fig:Survey-results} and \ref{fig:Money-plot}. It is clear that the new option AGNi is strongly favoured by the scientific community and would be rapidly picked up by authors and then by journals. Therefore, we have no concerns that this form of acronym for the plural would not survive in the literature.

\begin{table}
\caption{Results of the anonymous survey of acronym choice.}
\label{tab:Survey-results}
\begin{tabular}{rrrrr}
\textbf{AGN} & \hspace{1em} & \textbf{AGNs} & \hspace{1em} & \textbf{AGNi} \\ \hline
5 & & 5 & & 22
\end{tabular}
\end{table}

\section{Conclusion and future work}

In this short note, we have introduced a new standard acronym for Active Galactic Nuclei, namely AGNi. This acronym is superior in multiple ways, connecting to the Latin roots and cosmological theology, and also opens up linguistic opportunities in the discussion of AGNi, as well as new naming conventions for surveys or collaborations.

Going beyond this work, we can apply the same logic to all astronomy acronyms and initialisms that derive from Latin roots. Examples of this include Transient Lunar Phenomena (TLPa rather than TLPs), and the InterStellar Media \textit{etc.}~(ISMa, IGMa, ICMa rather than ISMs \textit{etc.}). These forms could be included in longer acronyms, such as Clumpy Hydrogen-$\alpha$-Rich InterStellar Media (CH$\alpha$RISMa) and ENergetic InterGalactic Media (ENIGMa).

\section*{Acknowledgements}

The authors would like to thank Lisa Kelsey, Emily Wickens and Michael Williams for informative discussions. We also thank Daniel Ballard, Molly Burkmar, Joe Callow, Elena Colangeli, Nathan Cruickshank, Fox Davidson, Wolfgang Enzi, Nathan Findlay, Bartolomeo Fiorini, Kieran Graham, Susanna Green, Rafaela Gsponer, Rhiannon Harries, Sophie Hoyland, Joseph Jackson, Shahab Joudaki, Lisa Kelsey, Sravan Kumar, Ricardo Landim, Konstantin Leyde, Tian Li, Sophie Newman, Daniela Saadeh, Ana Sainz De Murieta Martinez-Zubiri, Ashim Sen Gupta, Neel Shah, Sergi Sirera Lahoz, Han Wang, Luke Weisenbach, Arthur Whyley, Emily Wickens, Michael Williams, and the PhD students and postdocs at the Institute of Cosmology \& Gravitation for contributing to the anonymous survey.

\section*{Data Availability}

For the purpose of open access, the authors have applied a Creative Commons Attribution (CC BY) licence to any Author Accepted Manuscript version arising. Supporting research data are available on reasonable request from the corresponding author, Andrew Gow.



\bibliographystyle{mnras}
\bibliography{AGNi_paper} 





\bsp	
\label{lastpage}
\end{document}